\documentclass[usenatbib]{mn2e}
\input psfig

\newcommand{\araa}{ARA\&A}   % Annual Review of Astronomy and Astrophys.
\newcommand{\aj}{AJ}         % Astronomical Journal
\newcommand{\aaa}{A\&A}      % Astronomy and Astrophysics
\newcommand{\aas}{A\&AS}     % Astronomy and Astrophys. Supplement Series
\newcommand{\apj}{ApJ}       % Astrophysical Journal
\newcommand{\apjs}{ApJS}     % Astrophysical Journal Supplement Series
\newcommand{\apjl}{ApJ}      % Astrophysical Journal Letters
\newcommand{\mnras}{MNRAS}   % Monthly Notices of the Roy. Astron. Society
\newcommand{\pasp}{PASP}     % Publ. of the Astron. Society of the Pacific
\newcommand{\multi}{\multicolumn}

\def\picplace#1{\vbox{\hrule\@height 0.4pt\@width\hsize
\hbox to\hsize{\vrule\@width 0.4pt\@height#1\hfil
\vrule\@width 0.4pt\@height#1}\hrule\@height 0.4pt\@width\hsize}}
\def\squareforqed{\hbox{\rlap{$\sqcap$}$\sqcup$}}
\def\sq{\ifmmode\squareforqed\else{\unskip\nobreak\hfil
\penalty50\hskip1em\null\nobreak\hfil\squareforqed
\parfillskip=0pt\finalhyphendemerits=0\endgraf}\fi}

\def\la{\mathrel{\mathchoice {\vcenter{\offinterlineskip\halign{\hfil
$\displaystyle##$\hfil\cr<\cr\sim\cr}}}
{\vcenter{\offinterlineskip\halign{\hfil$\textstyle##$\hfil\cr
<\cr\sim\cr}}}
{\vcenter{\offinterlineskip\halign{\hfil$\scriptstyle##$\hfil\cr
<\cr\sim\cr}}}
{\vcenter{\offinterlineskip\halign{\hfil$\scriptscriptstyle##$\hfil\cr
<\cr\sim\cr}}}}}
\def\ga{\mathrel{\mathchoice {\vcenter{\offinterlineskip\halign{\hfil
$\displaystyle##$\hfil\cr>\cr\sim\cr}}}
{\vcenter{\offinterlineskip\halign{\hfil$\textstyle##$\hfil\cr
>\cr\sim\cr}}}
{\vcenter{\offinterlineskip\halign{\hfil$\scriptstyle##$\hfil\cr
>\cr\sim\cr}}}
{\vcenter{\offinterlineskip\halign{\hfil$\scriptscriptstyle##$\hfil\cr
>\cr\sim\cr}}}}}

\def\utw{\smash{\rlap{\lower5pt\hbox{$\sim$}}}}
\def\udtw{\smash{\rlap{\lower6pt\hbox{$\approx$}}}}

\def\farcm{\hbox{$.\mkern-4mu^\prime$}}
\def\farcs{\hbox{$.\!\!^{\prime\prime}$}}

\def\diameter{{\ifmmode\mathchoice
{\ooalign{\hfil\hbox{$\displaystyle/$}\hfil\crcr
{\hbox{$\displaystyle\mathchar"20D$}}}}
{\ooalign{\hfil\hbox{$\textstyle/$}\hfil\crcr
{\hbox{$\textstyle\mathchar"20D$}}}}
{\ooalign{\hfil\hbox{$\scriptstyle/$}\hfil\crcr
{\hbox{$\scriptstyle\mathchar"20D$}}}}
{\ooalign{\hfil\hbox{$\scriptscriptstyle/$}\hfil\crcr
{\hbox{$\scriptscriptstyle\mathchar"20D$}}}}
\else{\ooalign{\hfil/\hfil\crcr\mathhexbox20D}}%
\fi}}

\def\bbbc{{\mathchoice {\setbox0=\hbox{$\displaystyle\rm C$}\hbox{\hbox
to0pt{\kern0.4\wd0\vrule height0.9\ht0\hss}\box0}}
{\setbox0=\hbox{$\textstyle\rm C$}\hbox{\hbox
to0pt{\kern0.4\wd0\vrule height0.9\ht0\hss}\box0}}
{\setbox0=\hbox{$\scriptstyle\rm C$}\hbox{\hbox
to0pt{\kern0.4\wd0\vrule height0.9\ht0\hss}\box0}}
{\setbox0=\hbox{$\scriptscriptstyle\rm C$}\hbox{\hbox
to0pt{\kern0.4\wd0\vrule height0.9\ht0\hss}\box0}}}}
\def\bbbq{{\mathchoice {\setbox0=\hbox{$\displaystyle\rm
Q$}\hbox{\raise
0.15\ht0\hbox to0pt{\kern0.4\wd0\vrule height0.8\ht0\hss}\box0}}
{\setbox0=\hbox{$\textstyle\rm Q$}\hbox{\raise
0.15\ht0\hbox to0pt{\kern0.4\wd0\vrule height0.8\ht0\hss}\box0}}
{\setbox0=\hbox{$\scriptstyle\rm Q$}\hbox{\raise
0.15\ht0\hbox to0pt{\kern0.4\wd0\vrule height0.7\ht0\hss}\box0}}
{\setbox0=\hbox{$\scriptscriptstyle\rm Q$}\hbox{\raise
0.15\ht0\hbox to0pt{\kern0.4\wd0\vrule height0.7\ht0\hss}\box0}}}}
\def\bbbt{{\mathchoice {\setbox0=\hbox{$\displaystyle\rm
T$}\hbox{\hbox to0pt{\kern0.3\wd0\vrule height0.9\ht0\hss}\box0}}
{\setbox0=\hbox{$\textstyle\rm T$}\hbox{\hbox
to0pt{\kern0.3\wd0\vrule height0.9\ht0\hss}\box0}}
{\setbox0=\hbox{$\scriptstyle\rm T$}\hbox{\hbox
to0pt{\kern0.3\wd0\vrule height0.9\ht0\hss}\box0}}
{\setbox0=\hbox{$\scriptscriptstyle\rm T$}\hbox{\hbox
to0pt{\kern0.3\wd0\vrule height0.9\ht0\hss}\box0}}}}
\def\bbbs{{\mathchoice
{\setbox0=\hbox{$\displaystyle     \rm S$}\hbox{\raise0.5\ht0\hbox
to0pt{\kern0.35\wd0\vrule height0.45\ht0\hss}\hbox
to0pt{\kern0.55\wd0\vrule height0.5\ht0\hss}\box0}}
{\setbox0=\hbox{$\textstyle        \rm S$}\hbox{\raise0.5\ht0\hbox
to0pt{\kern0.35\wd0\vrule height0.45\ht0\hss}\hbox
to0pt{\kern0.55\wd0\vrule height0.5\ht0\hss}\box0}}
{\setbox0=\hbox{$\scriptstyle      \rm S$}\hbox{\raise0.5\ht0\hbox
to0pt{\kern0.35\wd0\vrule height0.45\ht0\hss}\raise0.05\ht0\hbox
to0pt{\kern0.5\wd0\vrule height0.45\ht0\hss}\box0}}
{\setbox0=\hbox{$\scriptscriptstyle\rm S$}\hbox{\raise0.5\ht0\hbox
to0pt{\kern0.4\wd0\vrule height0.45\ht0\hss}\raise0.05\ht0\hbox
to0pt{\kern0.55\wd0\vrule height0.45\ht0\hss}\box0}}}}
\def\bbbz{{\mathchoice {\hbox{$\sf\textstyle Z\kern-0.4em Z$}}
{\hbox{$\sf\textstyle Z\kern-0.4em Z$}}
{\hbox{$\sf\scriptstyle Z\kern-0.3em Z$}}
{\hbox{$\sf\scriptscriptstyle Z\kern-0.2em Z$}}}}

\newcommand{\VR}{$V\!-\!R$}
\newcommand{\VI}{$V\!-\!I$}

\newcommand{\lhun}{\llap{1}00}
\begin{document}

\title[{\it VRI\/} imaging photometry of globular clusters in the
  Magellanic clouds]{Integrated-light {\it VRI\/} imaging photometry
  of globular clusters in the Magellanic clouds}  

\author[P.\ Goudfrooij et al.]{Paul Goudfrooij,$^{1}$\thanks{E-mail (internet):
  goudfroo@stsci.edu}
Diane Gilmore,$^{1}$ 
Markus Kissler-Patig,$^{2}$\thanks{Visiting Astronomer, Cerro Tololo
  Inter-American Observatory. CTIO is operated by the Association of
  Universities for Research in Astronomy (AURA), Inc., under contract
  to the U.\,S. National Science Foundation.}and Claudia
  Maraston$^{3}$ \smallskip 
\\ 
$^1$\,Space Telescope Science Institute, 3700 San Martin Drive,
 Baltimore, MD 21218, U.S.A. \\
$^2$\,European Southern Observatory, Karl-Schwarzschild-Strasse 2, 
 D-85748 Garching bei M\"unchen, Germany \\ 
$^3$\,University of Oxford, Astrophysics, Denys Wilkinson Building,
 Keble Road, Oxford OX1 3RH, United Kingdom 
}

\date{Accepted 2006 March 14. Received 2006 March 13; in original form
  2006 February 14}  

\maketitle

\begin{abstract}
We present accurate integrated-light photometry in Johnson/Cousins $V$,
$R$, and $I$ for a sample of 28 globular clusters in the Magellanic
Clouds. The majority of the clusters in our sample have reliable age and
metallicity estimates available in the literature. The sample
encompasses ages between 50 Myr and 7 Gyr, and metallicities ([Fe/H])
between $-1.5$ and $0.0$ dex. The sample is dominated by 
clusters of ages between roughly 0.5 and 2 Gyr, an age range
during which the bolometric luminosity of simple stellar populations
is dominated by evolved red giant branch stars and thermally pulsing
asymptotic giant branch (TP-AGB) stars whose theoretical colours are
rather uncertain.  The {\it VRI\/} colours presented in this paper
have been used to calibrate stellar population synthesis model
predictions. 

\end{abstract}


%----------------------------- intro section ------------------------------

\section{Introduction}
\label{s:intro}

One of the predictions of stellar evolution theory is that evolved red
giant stars dominate the bolometric luminosity of a simple stellar
population (SSP) after a few hundred million years.  
According to predictions based on canonical stellar models
\citep{renbuz86}, the first (and sudden) appearance of a prominent red
stellar sequence is that of bright Asymptotic Giant Branch (AGB) stars
followed by the development of the Red Giant Branch (RGB), both
occurring before an age of 1 Gyr. The onset of the RGB is predicted to
occur at an age of about 0.6 Gyr \citep{renbuz86} and observational
efforts have  verified this prediction
\citep[e.g.,][]{ferr+95,ferr+04}. However, theoretical modelling of the AGB
phase is rather complicated. For example, strong and poorly
constrained mass loss and envelope burning affect stellar evolution
through the AGB phase
\citep[e.g.,][]{iberen83,waggro98,girber98,marigo01}. These phenomena
hamper a straightforward theoretical prediction of lifetimes and
effective temperatures during the thermally-pulsing part of the AGB
phase (the so-called TP-AGB phase). Hence, an empirical calibration of
the colours of SSPs during this age range is important \citep[see
also][]{renz92,mara98}.  

The globular cluster system of the Magellanic Clouds provides a unique
opportunity to study the influence of the AGB phase to the spectral
energy distribution of SSPs as a function of age and chemical
composition. Globular clusters in the Magellanic Clouds cover a wide
range in age and metallicity, and clusters with ages between 0.1 and 2
Gyr are present in significant numbers
\citep*[e.g.,][]{sear+80,elsfal85,sagpan89,frog+90,gira+95}.
\citet{mara98} presented SSP models whose calibration matches the
energetics and colours of globular clusters in the Magellanic
Clouds. The onset of the TP-AGB phase occurs at $\sim$\,0.2 Gyr and
lasts until $\sim$\,2 Gyr. During this age range, TP-AGB stars
dominate the near-IR and even the bolometric light of SSPs
\citep{mara98}. 

The age range of 0.2\,$-$\,2 Gyr, during which AGB stars are
expected to dominate the light, is of strong interest and significance
to several popular topics in present-day astronomy and  
astrophysics. For example, it is similar to the crossing time of
galaxy-sized stellar systems, which is a measure of the formation
timescale of a galaxy. Modern observational facilities like the {\it Hubble
  Space Telescope (HST)}, the {\it Spitzer Space Telescope\/} and
10-m-class ground-based telescopes have been able to identify various
populations of faint galaxies at high redshift ($z \ga 1$), for which
the accuracy of age determination is relevant to testing the
predictions of various galaxy formation scenarios \citep[see 
also][]{mara04}. This age range is 
also very relevant to the identification and study of post-starburst
galaxies and merger remnants \citep[e.g.,][and references
therein]{whit+93,schw+96,mara+01,goud+01a,goud+01b}. 

The models by \citet{mara98} were calibrated in the $U$, $B$, $V$,
$J$, $H$, and $K$ passbands. In order to allow for a wider
calibration of SSP models, we have obtained integrated-light
photometry in the $V$, $R_{\rm KC}$, and $I_{\rm KC}$ bands of 28
clusters in the Large and Small Magellanic Clouds (hereafter LMC and
SMC). We use digital CCD imaging and employ a range of aperture sizes
in order to evaluate the aperture size dependence of cluster colours
and their uncertainties. In contrast, virtually all previous studies
used single-aperture photometry or photographic plates. 

Our photometry has already been used by
\citet{mara05} for calibrating her SSP models in the full wavelength
range from $U$ to $K$. The influence of TP-AGB stars are found to be
significant in both the $R$ and $I$ passbands \citep[cf.\ Fig.~19
in][]{mara05}. In this paper we present the full data set. 

To our knowledge, this is the first study to present integrated magnitudes
in the $R$-- and $I$ bands for {\it any\/} LMC globular cluster and
for several SMC globular clusters as well. Among the
Johnson-Kron-Cousins passbands in the optical, $R$ and $I$ are most
affected by the AGB phase of stellar evolution. Hence, a calibration
of these passbands are important for proper identification of faint
young galaxies at cosmological redshifts ($1 \la z \la 6$) through
photometric redshift and spectral energy distribution studies in the
near-IR and mid-IR.  

\begin{table*}
\caption[]{General properties of globular clusters in our sample.}
\label{t:GCs}
\begin{tabular*}{\textwidth}{@{\extracolsep{\fill}}lccclccccc@{}} \hline \hline
\multi{10}{c}{~~} \\ [-1.8ex]  
\multicolumn{1}{l}{Cluster} & \multicolumn{1}{c}{RA} & \multicolumn{1}{c}{DEC} & SWB & 
 \multicolumn{1}{c}{Log (Age)} & Age & [Fe/H] & [Fe/H] & $A_V$ & $\sigma(A_V)$ \\
\multicolumn{1}{l}{~~~$\:$ID} & J2000 & J2000 & Type & \multicolumn{1}{c}{[yr]} & Ref. & 
 [dex] & Ref. & [mag] & [mag] \\
\multicolumn{1}{l}{~~~\,(1)} & (2) & (3) & (4) & \multicolumn{1}{c}{(5)} & (6) & (7) & (8) & (9) & (10) \\ [0.5ex] \hline
\multi{10}{c}{~~} \\ [-1.7ex] 
\multicolumn{10}{c}{SMC Clusters} \\ [0.5ex] \hline
\multi{10}{c}{~~} \\ [-1.4ex] 
Kron~3   & 00 24 44 & $-$72 47 20 & VI--VII & $9.78^{+0.09}_{-0.11}$ & 1   & $-1.16 \pm 0.09$ & 1   & 0.18\rlap{$^*$} & 0.02 \\ [0.5ex] 
NGC~152  & 00 32 56 & $-$73 06 57 & IV     &  $9.15^{+0.06}_{-0.07}$ & 2   & $-0.94 \pm 0.15$ & 2   & 0.19 & 0.02 \\ [0.5ex] 
NGC~265  & 00 47 12 & $-$73 28 38 & III    &  $8.30$                 & 3   & .....            &     & 0.20 & 0.02 \\ [0.5ex] 
NGC~339  & 00 57 49 & $-$74 28 00 & VII    &  $9.80^{+0.08}_{-0.10}$ & 4   & $-1.50 \pm 0.14$ & 4   & 0.18 & 0.02 \\ [0.5ex] 
NGC~411  & 01 07 56 & $-$71 46 05 & V--VI  &  $9.15^{+0.06}_{-0.07}$ & 2,4 & $-0.68 \pm 0.07$ & 2,4 & 0.17 & 0.02 \\ [0.5ex] 
NGC~419  & 01 08 29 & $-$72 53 12 & V      &  $9.08^{+0.15}_{-0.23}$ & 5   & $-0.7  \pm 0.3$  & 5   & 0.32 & 0.02 \\ [0.5ex] \hline
\multi{10}{c}{~~} \\ [-1.7ex] 
\multicolumn{10}{c}{LMC Clusters} \\ [0.5ex] \hline
\multi{10}{c}{~~} \\ [-1.4ex] 
NGC~1644 & 04 37 39 & $-$66 11 58 & V     &  $8.94$                 & 6  & .....            &      & 0.28\rlap{$^*$} & 0.02 \\  [0.5ex] 
NGC~1651 & 04 37 32 & $-$70 35 06 & V     &  $9.30^{+0.08}_{-0.10}$ & 7  & $-0.37 \pm 0.20$ &  12  & 0.35 & 0.05 \\  [0.5ex] 
NGC~1751 & 04 54 12 & $-$69 48.24 & VI    &  $9.18$                 & 6  & $-0.18 \pm 0.20$ &  12  & 0.51 & 0.02 \\ [0.5ex] 
NGC~1755 & 04 55 14 & $-$68 12 17 & II    &  $7.98$                 & 6  & .....            &      & 0.55 & 0.02 \\ [0.5ex] 
NGC~1783 & 04 59 08 & $-$65 59 20 & V     &  $9.11$                 & 6  & $-0.45 \pm 0.03$ &  13  & 0.31 & 0.02 \\ [0.5ex] 
NGC~1806 & 05 02 11 & $-$67 59 20 & V     &  $8.70$                 & 8  & $-0.23 \pm 0.20$ &  12  & 0.25 & 0.04 \\ [0.5ex]
NGC~1831 & 05 06 16 & $-$64 55 06 & IV\,A &  $8.50^{+0.30}_{-0.30}$ & 9  & $+0.01 \pm 0.20$ &  12  & 0.39\rlap{$^*$} & 0.02 \\ [0.5ex] 
NGC~1846 & 05 07 35 & $-$67 27 39 & VI    &  $9.08$                 & 6  & $-0.70 \pm 0.20$ &  12  & 0.45 & 0.03 \\ [0.5ex] 
NGC~1866 & 05 13 39 & $-$65 27 54 & III   &  $8.12^{+0.30}_{-0.30}$ & 9  & $-0.50 \pm 0.10$ &  14  & 0.26 & 0.03 \\ [0.5ex] 
NGC~1868 & 05 14 36 & $-$63 57 18 & IV\,A &  $8.74^{+0.30}_{-0.30}$ & 9  & $-0.50 \pm 0.20$ &  12  & 0.39\rlap{$^*$} & 0.02 \\ [0.5ex] 
NGC~1978 & 05 28 45 & $-$66 14 12 & VI    &  $9.30^{+0.05}_{-0.05}$ & 7  & $-0.42 \pm 0.20$ &  12  & 0.55 & 0.03 \\ [0.5ex] 
NGC~1987 & 05 27 17 & $-$70 44 06 & IV\,B &  $8.79$                 & 6  & .....            &      & 0.33 & 0.03 \\ [0.5ex]
NGC~2058 & 05 36 54 & $-$70 09 44 & III   &  $8.06$                 & 6  & .....            &      & 0.24 & 0.02 \\ [0.5ex] 
NGC~2134 & 05 51 56 & $-$71 05 51 & III   &  $8.28$                 & 6  & $-1.0$           &  15  & 0.49 & 0.02 \\ [0.5ex] 
NGC~2136 & 05 53 17 & $-$69 31 42 & III   &  $8.00^{+0.10}_{-0.10}$ & 8  & $-0.55 \pm 0.23$ &  8   & 0.46 & 0.02 \\ [0.5ex] 
NGC~2154 & 05 57 38 & $-$67 15 43 & V     &  $9.01$                 & 6  & $-0.56 \pm 0.20$ &  12  & 0.34 & 0.02 \\ [0.5ex] 
NGC~2155 & 05 58 33 & $-$65 28 36 & VI    &  $9.51^{+0.06}_{-0.07}$ & 10 & $-0.55 \pm 0.20$ &  12  & 0.35 & 0.03 \\ [0.5ex] 
NGC~2162 & 06 00 31 & $-$63 43 18 & V     &  $9.11^{+0.12}_{-0.16}$ & 7  & $-0.23 \pm 0.20$ &  12  & 0.39 & 0.02 \\ [0.5ex]
NGC~2164 & 05 58 54 & $-$68 31 06 & II    &  $7.70^{+0.20}_{-0.20}$ & 11 & $-0.60 \pm 0.20$ &  16  & 0.41 & 0.07 \\ [0.5ex] 
NGC~2173 & 05 57 58 & $-$72 58 42 & VI    &  $9.33^{+0.07}_{-0.09}$ & 7  & $-0.24 \pm 0.20$ &  12  & 0.39\rlap{$^*$} & 0.02 \\ [0.5ex] 
NGC~2213 & 06 10 42 & $-$71 31 42 & V     &  $9.20^{+0.10}_{-0.12}$ & 7  & $-0.01 \pm 0.20$ &  12  & 0.44 & 0.02 \\ [0.5ex] 
NGC~2231 & 06 20 44 & $-$67 31 06 & V     &  $9.18^{+0.10}_{-0.13}$ & 7  & $-0.67 \pm 0.20$ &  12  & 0.39 & 0.02 \\ [0.5ex] \hline 
\multi{10}{c}{~~} \\ [-1.2ex] 
\end{tabular*}
\parbox{\textwidth}{
{\sl Notes to Table \ref{t:GCs}}.~~Column (1): Cluster ID. Columns (2) and (3): 
Right Ascension (given as hours, 
minutes, and seconds) and Declination (given as degrees, arcminutes, and
arcseconds)) in J2000.0 equinox, both taken from Welch \citeyearpar{welch91}
for the SMC clusters and from Bica et al.\ \citeyearpar{bica+99} for the LMC
clusters. Column (4): SWB type from Frogel et al.\ 
\citeyearpar{frog+90} or Bica et al.\ \citeyearpar{bica+96}. Column (5): Logarithm
of age from literature. Column (6): Age reference code (see below). Column (7):
[Fe/H] from literature. Column (8): [Fe/H] reference code (see below). Columns
(9) and (10): Assumed foreground extinction in $V$-band and its uncertainty,
derived from data in Zaritsky et al.\ \citeyearpar{zar+02,zar+04}, made
available through http://ngala.as.arizona.edu/dennis/lmcdata.html. Entries with asterisk superscript were derived using \citet{schl+98}. See text in
Sect.~\ref{s:extin}. \\
\hspace*{10pt} {\sc References:} (1) Mighell et al.\ \citeyearpar{migh+98}; 
(2) Crowl et al.\ \citeyearpar{crow+01}; (3) Maraston \citeyearpar{mara05}; (4) 
Da Costa \& Hatzidimitriou \citeyearpar{dachat98}; (5) Durand et al.\ 
\citeyearpar{dura+84};(6) Girardi \& Bertelli \citeyearpar{girber98}; 
(7) Geisler et al.\ \citeyearpar{geis+97}; (8) Dirsch et al.\
\citeyearpar{dirs+00}; (9) Elson \& Fall \citeyearpar{elsfal88}; (10) Rich et al.\
\citeyearpar{rich+01}; (11)  Elson \citeyearpar{elson91}; (12) Olszewski et al.\
\citeyearpar{olsz+91}; (13)  Cohen \citeyearpar{cohen82}; (14) Hill et al.\
\citeyearpar{hill+00}; (15) Hodge \citeyearpar{hodge84}; (16) Schommer \& Geisler
\citeyearpar{schgei88}.   
} 
\end{table*}

\subsection{Sample selection}
\label{s:sample}

Taking the sample of bright globular clusters studied by \citet{frog+90} as a
starting point, we selected a sample of star clusters with Searle, Wilkinson
\& Bagnuolo (\citeyear{sear+80}, hereafter SWB) types between III and VI with the
purpose to select clusters in the age range of $0.3 \la t \la 2$ Gyr (e.g.,
Frogel et al.\ \citeyear{frog+90}), which is relevant to constraining the
influence of AGB stars on the integrated colours of stellar populations. Care
was taken to select a suitably high and approximately equal number of
clusters in each SWB type between III and VI. As individual clusters have
masses of a few $10^4 M_{\odot}$ and lifetimes of
luminous AGB stars are quite short ($\la 10^7$ yr), stochastic fluctuations
strongly affect the observed number of AGB stars per cluster
\citep[see also][]{mara98,bc03}. Hence, the selection of several
clusters in each SWB type allows one to set empirical constraints on
such stochastic effects as a function of age.  
Relevant global parameters of the star clusters in our sample are listed in
Table~\ref{t:GCs}. 

\section{Observations and Basic Data Reduction}

Observations were performed during two runs at the 0.9-m telescope of
Cerro Tololo Interamerican Observatory (CTIO) in f/13.5 mode using the
CCD imager and a dedicated 2048x2048 SITe CCD, which was employed
in gain setting 2 (1.5 e$^-$/ADU), yielding a read noise of 3.6 e$^-$
per pixel. With a plate scale of 0\farcs40 pixel$^{-1}$, the field of
view is 13\farcm6\,$\times$\,13\farcm6. Images were taken through
Johnson $V$ and Kron-Cousins $R$ and $I$ filters. Other details on the
observation runs are listed in Table~\ref{t:obsruns}. 

Bias and dome flat images as well as images of the sky during twilight
(``sky flats'') were acquired daily and used for basic CCD
calibration using the {\sc ccdproc} package of IRAF. During
acquisition of the sky flat frames, care was taken to acquire several
images per filter with 10-arcsecond offsets of the telescope between
images so that any stars in the field as well as bad pixels and cosmic
ray hits could be eliminated during image combination and smooth
illumination correction frames could be created. Standard star fields
from Landolt \citeyearpar{land92} with an appropriate range of intrinsic colours
were observed several times per night to derive photometric
transformation equations and extinction corrections. Typically, a
total of three or four standard star fields were observed two or three
times each per night. The log of Magellanic cloud star cluster observations
is listed in Table~\ref{t:obslog}. 

Each target globular cluster was centered roughly 2 arcmin away from
the centre of the CCD in order to maximize the area usable for 
the determination of the background level (see Sect.\ \ref{s:photom}
below), yet still cover the full spatial extent of the target
clusters. Multiple exposures were taken through each filter. The
telescope was offset by 5 arcsec between each exposure,  which
facilitated the elimination of cosmic ray hits and hot pixels during
image combination. The latter was done using {\sc iraf} task {\sc
  imcombine} using the {\sc crreject} option after having aligned
the subimages in each filter to a common coordinate system using
centroids of relatively isolated stars in the field. The alignment
accuracy was always better than 0.05 pixels in each axis.    

The weather during the January 2002 run was photometric, and the
photometric zeropoint calibrations stayed constant throughout that run to
within 0.01 mag.  
The December 2004 run, which was performed in service mode operated
by the SMARTS\footnote{Small and Moderate Aperture Research Telescope
  System, see http://www.astro.yale.edu/smarts} consortium, suffered
from non-photometric weather. The photometric transformation
equations for these images were derived by comparing observed count
rates of several tens of stars in the field of view of each cluster image
with $V$ and $I$-band photometry published by Zaritsky et al.\
\citeyearpar{zar+02,zar+04} and $R$-band photometry by Massey
\citeyearpar{massey02} as follows. We first selected 
stars from the Zaritsky et al.\ and Massey catalogs with celestial
positions between 3 and 10 arcmin away from the cluster centre
positions. The latter were taken from Bica et al.\ \citeyearpar{bica+99}
for the LMC clusters and Welch \citeyearpar{welch91} for the SMC
clusters. (The inner radius threshold of 3 arcmin was put in place to
avoid the crowded areas near the cluster centres.) The astrometric
zeropoint offset between the Zaritsky et al.\ and Massey catalogs and
the world coordinate system of our images was determined for each
cluster by identifying (by eye, using the ESO {\sc skycat} tool) a
star in those catalogs located near the centre of our image and
measuring the pixel location of the centroid of that star on our
image. Estimated pixel positions of all other stars (selected as
mentioned above) around the cluster in question are then
calculated. This list of pixel positions is subsequently used as input
to the {\sc daophot-ii} package 
\citep{stet87} as implemented within {\sc iraf} to calculate
aperture photometry for those stars on our images. Comparing the input
list of pixel positions with positions in the final list of
instrumental magnitudes of those stars shows that the positional
agreement was better than 1.5 pixels RMS in all cases, good enough to
avoid confusion due to crowding. Aperture corrections are derived
separately for each filter and each cluster by using curve-of-growth
measurements for several isolated stars in the field.  Finally,
photometric transformation solutions were derived separately for each
cluster observed during the Dec 2004 run. These solutions involved a
zero-point term and one linear colour term (using \VI\ for $V$ and $I$
and \VR\ for $R$). The RMS uncertainty of these photometric solutions
ranges between 0.01 and 0.04 mag, which is smaller than other
sources of error in the measurement of integrated cluster colours (see
below).  

%
%------------------------------ TABLE 2 ---------------------------------
%
\begin{table}
\caption[ ]{Observing runs at the CTIO 0.9-m telescope.}
\label{t:obsruns}
\begin{tabular*}{8.4cm}{@{\extracolsep{\fill}}lll} \hline \hline
\multi{3}{c}{~~} \\ [-1.8ex]  
Run   & Jan 02 & Dec 04 \\ [0.5ex] \hline 
\multi{3}{c}{~~} \\ [-1.2ex] 
Allocation & NOAO & SMARTS \\
Dates & Jan 10--13, 2002 & Dec 9--11, 2004 \\
Observer & Kissler-Patig & Gomez$^1$ \\
Conditions        & Photometric & Variable clouds \\
\multi{3}{c}{~~} \\ [-1.8ex]  \hline 
\multi{3}{c}{~~} \\ [-1.8ex] 
\end{tabular*}

\noindent
{\small
$^1$ A. Gomez, SMARTS service observer. }
\end{table}

%------------------------------ TABLE 3 ---------------------------------
%
\begin{table}
\caption[ ]{Observing log}
\label{t:obslog}
\begin{tabular*}{8.4cm}{@{\extracolsep{\fill}}lrcccc@{}} \hline \hline
\multi{3}{c}{~~} \\ [-1.8ex]  
Object & \multicolumn{1}{c}{Observing} & \multicolumn{3}{c}{Exposure
  Time (s)} & $V$-band \\ 
       & \multicolumn{1}{c}{Date}      & $V$    & $R$    & $I$              
    & seeing$^*$  \\ [0.5ex] \hline 
\multi{3}{c}{~~} \\ [-1.2ex] 
  Kron~3 & Dec 10, 2004 & 3$\times$420 & 3$\times$240 & 3$\times$450 &  1.19 \\        
 NGC~152 & Jan 10, 2002 & 3$\times$500 & 3$\times$500 & 3$\times$600 &  1.60 \\        
 NGC~265 & Jan 13, 2002 & 3$\times$400 & 3$\times$250 & 3$\times$400 &  1.77 \\        
 NGC~339 & Dec 11, 2004 & 3$\times$420 & 3$\times$240 & 3$\times$450 &  1.31 \\        
 NGC~411 &  Dec 9, 2004 & 3$\times$420 & 3$\times$240 & 3$\times$450 &  1.25 \\        
 NGC~419 & Jan 12, 2002 & 3$\times$400 & 3$\times$250 & 3$\times$400 &  1.74 \\ [0.8ex]
NGC~1644 & Jan 10, 2002 & 3$\times$200 & 3$\times$200 & 3$\times$400 &  1.47 \\        
NGC~1651 & Dec 11, 2004 & 3$\times$210 & 3$\times$150 & 3$\times$240 &  1.24 \\        
NGC~1751 & Jan 13, 2002 & 3$\times$180 & 3$\times$140 & 3$\times$200 &  1.56 \\        
NGC~1755 & Jan 13, 2002 & 3$\times$180 & 3$\times$140 & 3$\times$200 &  1.55 \\        
NGC~1783 & Jan 12, 2002 & 3$\times$180 & 3$\times$140 & 3$\times$200 &  1.50 \\ [0.8ex]
NGC~1806 &  Dec 9, 2004 & 3$\times$210 & 3$\times$150 & 3$\times$240 &  1.33 \\        
NGC~1831 & Jan 12, 2002 & 3$\times$180 & 3$\times$140 & 3$\times$200 &  1.31 \\        
NGC~1846 & Jan 12, 2002 & 3$\times$180 & 3$\times$140 & 3$\times$200 &  1.54 \\        
NGC~1866 & Jan 10, 2002 & 3$\times$300 & 3$\times$250 & 3$\times$500 &  2.02 \\ 
NGC~1868 & Jan 13, 2002 & 3$\times$180 & 3$\times$140 & 3$\times$200 &  1.45 \\ [0.8ex]     
NGC~1978 & Jan 10, 2002 & 3$\times$200 & 3$\times$200 & 3$\times$400 &  1.71 \\        
NGC~1987 & Jan 10, 2002 & 3$\times$250 & 3$\times$200 & 3$\times$400 &  2.30 \\        
NGC~2058 & Jan 13, 2002 & 3$\times$180 & 3$\times$140 & 3$\times$200 &  2.12 \\        
NGC~2134 & Jan 12, 2002 & 3$\times$180 & 3$\times$140 & 3$\times$200 &  1.31 \\ 
NGC~2136 & Jan 12, 2002 & 3$\times$180 & 3$\times$140 & 3$\times$200 &  1.54 \\ [0.8ex]     
NGC~2154 & Jan 13, 2002 & 3$\times$180 & 3$\times$140 & 3$\times$200 &  1.35 \\        
NGC~2155 & Jan 12, 2002 & 3$\times$180 & 3$\times$140 & 3$\times$200 &  1.77 \\        
NGC~2162 & Jan 13, 2002 & 3$\times$180 & 3$\times$140 & 3$\times$200 &  1.40 \\        
NGC~2164 &  Dec 9, 2004 & 3$\times$210 & 3$\times$150 & 3$\times$240 &  1.23 \\ 
NGC~2173 & Jan 13, 2002 & 3$\times$180 & 3$\times$140 & 3$\times$200 &  1.25 \\ [0.8ex]
NGC~2213 & Jan 12, 2002 & 3$\times$180 & 3$\times$140 & 3$\times$200 &  1.92 \\
NGC~2231 & Jan 12, 2002 & 3$\times$180 & 3$\times$140 & 3$\times$200 &  1.82 \\
\multicolumn{3}{c}{~~} \\ [-1.8ex]  \hline 
\multicolumn{3}{c}{~~} \\ [-1.8ex] 
\multicolumn{6}{l}{\small $^*$\,Seeing FWHM values in arcsec.}
\end{tabular*}

\noindent
\end{table}

\section{Integrated-light cluster photometry}
\label{s:photom}

\subsection{Measurement technique}

The measurement of integrated magnitudes and colours of globular
clusters in the Magellanic clouds is complicated by several
factors. One problem is that of accurate centering of the measurement
aperture. Many of these clusters are superposed onto a relatively high
surface density of stars associated with the LMC or SMC, and some have
a rather irregular star distribution and/or are not particularly
symmetric due to the superposition of bright stars (be it supergiants or
AGB stars within the cluster itself, those associated with the body of the
LMC or SMC, or Galactic foreground stars). On the other hand, it should be
recognized that the use of CCD images in this context renders these
problems much less severe than they were for previous studies which used
single-channel photometers and diaphragms which were centered by eye or by
maximum throughput.  
After some experimentation, we decided to employ the following
centering method for each cluster. Using the point spread
function-fitting mode of the {\sc daophot-ii} package, we first removed
stars brighter than the magnitude of an O9 supergiant star ($M_V = -6.4$,
Schmidt-Kaler 1982) at the distance of the LMC or SMC (for this
purpose we adopt $(m\!-\!M)$ = 18.24 for the LMC and $(m\!-\!M)$ = 18.75 for the
SMC; Udalski 2000) from the $V$-band image. In order to determine a proper 
luminosity-weighted centre, we then convolved this image by a series of
circular Gaussians with standard deviations of 20 through 60 pixels (i.e., 8
through 24 arcsec), and the centroids of the clusters (and their
uncertainties) were measured for each, starting with an initial guess
estimated by eye. The final cluster centre was defined as the mean centroid
weighted by the inverse variances of the individual centroid fits. The
standard error of the cluster centre was taken to be the standard deviation of
the list of cluster centres as derived above. 

Since globular clusters vary significantly in size, we chose to
measure the cluster photometry using several aperture sizes, including
the (typically single) aperture size used by previous studies to allow
a simple comparison. The apertures are circular and the same for all
filter bands. 
The subtraction of ``background'' light not belonging to the target
clusters themselves (i.e., sky, Galactic foreground stars, and the diffuse
stellar population associated with the LMC or SMC at the location of
the target clusters) was performed as follows. We started out with the
images from which single stars brighter than the equivalent of $M_V =
-6.4$ were removed (see above). Areas affected by CCD saturation or
obvious scattered light from internal reflections due to the presence
of bright stars were also flagged and excluded from background level
measurements. The mean background level (and its uncertainty) was 
measured from the (remaining) area on the cluster images away from the
cluster centre by 3 arcmin (i.e., $\sim$\,40 pc) or more. In the cases
where more than one cluster is present on the image (e.g., NGC~1755,
NGC~2058, NGC~2065), we also excluded pixels within 3 arcmin from
those clusters. To arrive at the final cluster count rates in each
filter, this mean background level was multiplied by the number of
pixels in a given source aperture and subtracted from the summed
cluster counts within that aperture.   
Uncertainties for the final cluster magnitudes and colours were
estimated by remeasuring cluster count rates through apertures centred
on 4 distinct positions away by 3\,$\sigma$ from the derived cluster
centre and adding the formal uncertainty of the photometric
calibration in quadrature. The final magnitudes and colours of the target
clusters are listed for aperture radii of 30$''$, 45$''$, 60$''$, and
100$''$ in Table~\ref{t:photdata} (in the Appendix). Magnitudes and
colours for other radii can be provided by the first author on request. 

\subsection{Comparison with previous studies}
\label{s:compare}

\begin{figure}
%\centerline{\psfig{figure=us_vs_vdb.eps,width=6.5cm}}
\centerline{\psfig{figure=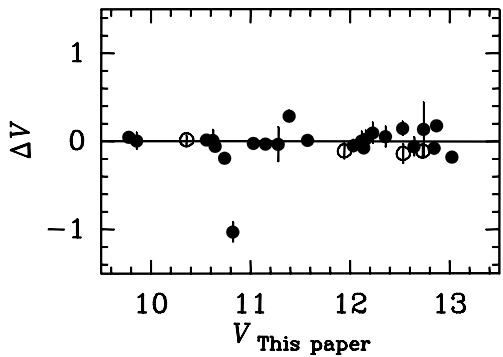,width=6.5cm}}
\caption{Comparison of our $V$ magnitudes with those from
  \citet{vdb81}. The difference $\Delta V$ is in the sense ``our
  values minus those of van den Bergh''. Filled circles represent data
  take during the Jan 2002 observing run, open circles represent data
  taken during the Dec 2004 run. The errorbars only represent our
  uncertainties, since the data in \citet{vdb81} do not contain
  photometric uncertainties. The solid line represents $\Delta V = 0$.} 
\label{f:us_vs_vdb}
\end{figure}

We obtain an external measure of the accuracy of our measurements by
comparing our results with previous studies. As we
found no prior integrated photometry in the $R$-- and $I$-bands for our
LMC clusters in the literature, we limit this comparison to one 
with $V$ magnitudes from the large compilation of integrated {\it
  UBV\/} photometry by \citet{vdb81}, which contains $V$-band data for all
clusters in our sample. This comparison is shown in  
Figure~\ref{f:us_vs_vdb}. We derived $V$ magnitudes for our data using
the aperture radii listed in Tables~2 and 3 of \citet{vdb81} for the 
purpose of this comparison. As van den Bergh did not provide
photometric uncertainties for the data he compiled, the error bars
plotted in Figure~\ref{f:us_vs_vdb} only reflect our uncertainties. In
any case, it can be seen that the data agree very well with each
other. The formal difference in $V$ magnitudes, $\Delta V \equiv
V_{\rm this\;paper} - V_{\rm van\;den\;Bergh} = -0.03 \pm 0.04$ where the
error estimate represents the mean error of the mean. 
There is one outlier, which is the young cluster NGC~2058, where we
measure a significantly brighter $V$ magnitude than van den
Bergh. NGC~2058 is located in a rather crowded region of the LMC with
significant variations in the surface brightness of the field
population outside the cluster. It is also surrounded by neighbouring
clusters. To understand this discrepancy, we investigated three
possibilities:\ 
{\it (i)\/} An unfortunate 
miscentering of the aperture used by the single-channel photometry
reported in \citet{vdb81}, which might have been the case if the
aperture was centered using a blue-sensitive eyepiece at the time
(single bright young stars located off the cluster centre might cause
such an effect; see Pessev et al.\ \citeyearpar{pess+06} for
illustrations of this effect in the near-IR). However, in order to
arrive at van den Bergh's $V$ magnitude, we find that one would have
to locate the aperture $\sim$\,32$''$ off the cluster centre, which is
half the diameter of the van den Bergh aperture. This seems unlikely,
but cannot be ruled out a priori. 
{\it (ii)\/} An unfortunate placement of the background aperture of
the measurement of NGC~2058 in van den Bergh's compilation onto a
neighbouring (e.g., faint) star cluster. However, we could only
account for an error of 0.45 mag that way (placing the background
aperture centered on the brightest neighbouring cluster on the CCD),
which is still 0.55 mag short of explaining the discrepancy. 
{\it (iii)\/} An unfortunate error in our photometric calculations. However,
we did double-check our measurements, and the $V$-band magnitudes of all
other clusters observed during the same night (for which the photometry was
derived in the exact same way as for NGC~2058) {\it are\/} consistent with
those listed in \citet{vdb81}. 
We conclude that in the absence of knowledge of the location of the
object and background apertures for NGC~2058 in the \citet{vdb81}
compilation, it is impossible to pinpoint the reason for this
discrepancy. For now, we suggest that it may be at least partly due to
aperture miscentering and/or unfortunate background aperture placement
as described above. 

\begin{figure*}
%\centerline{\psfig{figure=us_vs_rz.eps,width=13cm}}
\centerline{\psfig{figure=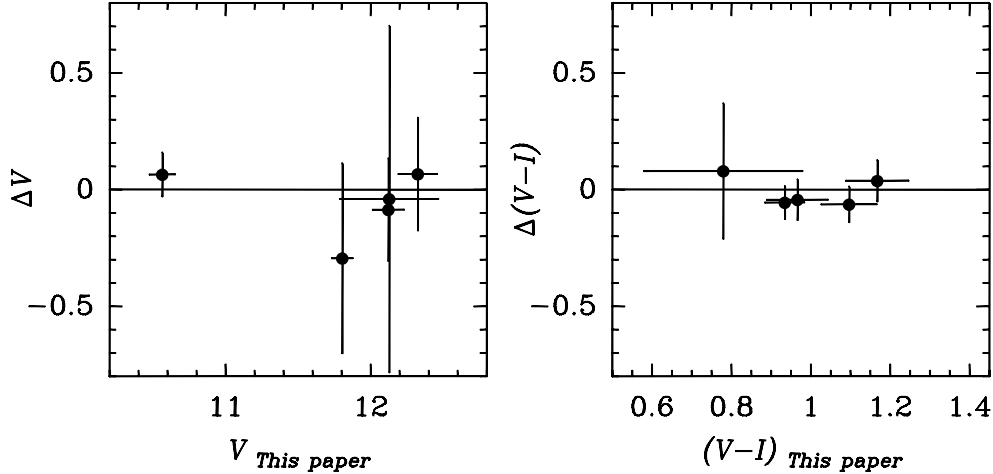,width=13cm}}
\caption{Comparison of our magnitudes and colours with those from
  \citet{rafzar05} for the clusters in common between the two studies
  (NGC~152, NGC~265, NGC~339, NGC~411, and NGC~419). $V$ magnitudes are
  compared in the left panel, and $V\!-\!I$ colours are compared in the right
  panel. The differences $\Delta V$ and $\Delta (V\!-\!I)$ are in the
  sense ``our values minus those of Rafelski \& Zaritsky''. The solid
  lines represent $\Delta V = 0$ (left panel) and $\Delta (V\!-\!I) =
  0$ (right panel).}  
\label{f:us_vs_rz}
\end{figure*}

For the SMC clusters in our sample, we compare our $V$ magnitudes and \VI\
colours with those measured by the recent CCD imaging study of
\citet[][hereafter RZ05]{rafzar05} in Fig.~\ref{f:us_vs_rz}. 
As RZ05 used cluster radii from Hill \& Zaritsky
\citeyearpar{hilzar06} for their {\it UBVI\/} measurements, we did so
too for this comparison.   
The agreement in both $V$ magnitudes and \VI\ colours is within the
uncertainties, which is encouraging since they used a similar
measurement method. The formal differences in $V$ magnitudes and \VI\
colours are 
$\Delta V  \equiv V_{\rm this\;paper} - V_{\rm RZ05} = -0.06 \pm
0.03$ and 
$\Delta (V\!-\!I) \equiv (V\!-\!I)_{\rm this\;paper} -
(V\!-\!I)_{\rm RZ05} = -0.01 \pm 0.01$, 
respectively, where the error estimates represent the mean error of
the mean. The average errors of the $\Delta V$ and $\Delta (V\!-\!I)$
values themselves are 0.34 mag and 0.12 mag, respectively.

\subsection{Extinction Corrections}
\label{s:extin}

Cluster extinction values were obtained from two independent studies:
Schlegel, Finkbeiner, \& Davis \citeyearpar{schl+98}, and Zaritsky et
al.\ \citeyearpar{zar+02,zar+04}. While the Schlegel et al.\
\citeyearpar{schl+98} measurements cover 
the whole sky and use direct measurements of dust emission, it has been
recognized that their extinction maps are rather uncertain in the inner
regions of the Magellanic Clouds due to the lack of adequate temperature
structure resolution by DIRBE and the fact that their extinction values might 
be systematically underestimated due to the possible presence of cold dust
invisible to IRAS.  With this in mind, our adopted   
extinction values were determined primarily from the 
measurements of Zaritsky et al.\ \citeyearpar{zar+02,zar+04} which are based
on stellar atmosphere model fits to {\it UBVI\/} photometry of
thousands of individual stars across the SMC and LMC. 
We chose to derive extinction values from the ``cool'' stars ($T_{\rm eff} <
10^4$ K) in the Zaritsky et al.\ surveys. This choice was made after
considering that {\it (i)\/} our objects are all significantly older than 10
Myr, and {\it (ii)\/} the $U$-band photometry from Zaritsky et al.\ is 
significantly less deep than their $B$, $V$, and $I$-band photometry,
rendering their stellar atmosphere model fitting procedure to be typically
more precise for cooler stars. The clusters were assigned an extinction
value that is the mean of the distribution of values returned by the
Zaritsky et al.\ method, while its uncertainty was assigned the mean
error of that mean value. We ended up using a search
radius of 5$'$ around the center coordinates of the target star
clusters, as the use of smaller search radii sometimes returned rather few
stars with extinction estimates. From experimentation with different values
of the search radius, we estimate the systematic uncertainty of the
$A_V$ extinction values listed in Table~\ref{t:GCs} to be within 0.02
mag. For those target clusters not covered by the Zaritsky et al.\ surveys 
(whose $A_V$ values are indicated with asterisk superscripts in Table~\ref{t:GCs}), 
we use extinction values from Schlegel et al.\ \citeyearpar{schl+98} and
correct them to the Zaritsky et al.\ extinction scale by adding the mean
difference of $A_V$ values for clusters for which both extinction values are
available. This mean difference was determined to be $A_{V,\: {\rm Zaritsky\;
 et\; al.}} - A_{V,\: {\rm Schlegel\; et\; al.}} = 0.02 \pm 0.01$. 
Uncertainties of extinction values derived from Schlegel et al.\
\citeyearpar{schl+98} were assigned the mean uncertainty of the extinction
values derived from the Zaritsky et al.\ surveys in the SMC or LMC, 
depending on the cluster in question, with the 0.01 mag of uncertainty
mentioned above added in quadrature. The conversion of $A_V$ values to
$A_R$ and $A_I$ was done using the formulae in Cardelli, Clayton, \& Mathis 
\citeyearpar{card+89} with $R_V$ = 2.7 and 3.4 for the SMC and LMC,
respectively \citep{gord+03}. We used reference wavelengths of 6407
\AA\ and 7982 \AA\ for the Kron-Cousins $R$ and $I$ bands,
respectively \citep{bess90}. The final extinction values are listed in
Table~\ref{t:GCs}. Dereddened cluster colours $(V\!-\!R)_0$ and
$(V\!-\!I)_0$ are listed in Table~\ref{t:photdata}.

\section{Concluding remarks}

We have presented accurate integrated-light photometry in Johnson/Cousins $V$,
$R$, and $I$ for a sample of 28 globular clusters in the Magellanic
Clouds, most of which have reliable age and metallicity estimates
available in the literature. The sample encompasses estimated ages between 50
Myr and 7 Gyr, and metallicities ([Fe/H]) between $-1.50$ and $+0.01$
dex. The sample is dominated by clusters of ages between roughly 0.5
and 2 Gyr, an age range during which the bolometric luminosity of
simple stellar populations is dominated by AGB stars whose
theoretical colours are rather uncertain. Hence, an 
empirical calibration of the colours of SSPs during this age range is
important. Indeed, the data presented here have recently been used by
\citet{mara05} to update the calibration of SSP models, finding that
the influence of AGB effects is significant in $R$ and $I$. To our
knowledge, this is the first study to present $R$-band photometry for
any cluster in our sample, and $I$-band photometry for any LMC cluster.

\paragraph*{Acknowledgments.} \ \\ 
This work is based on observations obtained at the Cerro Tololo
Interamerican Observatory, which is is operated by the Association of
Universities for Research in Astronomy (AURA), Inc., under contract
to the U.\,S. National Science Foundation. This research has made use of
the NASA/IPAC Extragalactic Database (NED) which is operated by the Jet
Propulsion Laboratory, California Institute of Technology, under contract
with the National Aeronautics and Space Administration. We thank CTIO
staff observer Arturo Gomez for his help in securing the observations of
the December 2004 run operated by the SMARTS consortium. We acknowledge
helpful discussions with Peter Pessev, and thank the anonymous referee
for a fast and thorough review of this paper. PG and DG also thank the
Director of STScI for funding part of this research through a
director's discretionary research grant.

\appendix

\section{Aperture photometry of 28 globular clusters in the Magellanic clouds}

\begin{table*}
\small
\caption[ ]{{\it VRI\/} Photometry of globular clusters in the Magellanic clouds.} 
\label{t:photdata}
\begin{tabular*}{15cm}{@{\extracolsep{\fill}}lcrrrrr@{}} \hline \hline
\multicolumn{7}{c}{~~} \\ [-1.8ex]
Object & Aperture & \multicolumn{1}{c}{$V$} & \multicolumn{1}{c}{$V\!-\!R$} &
       \multicolumn{1}{c}{$V\!-\!I$} & \multicolumn{1}{c}{$(V\!-\!R)_0$} & 
       \multicolumn{1}{c}{$(V\!-\!I)_0$} \\
       & Radius\rlap{$^*$} & \multicolumn{1}{c}{[mag]} & \multicolumn{1}{c}{[mag]} & 
  \multicolumn{1}{c}{[mag]} & \multicolumn{1}{c}{[mag]} & 
  \multicolumn{1}{c}{[mag]} \\ [0.5ex] \hline
\multicolumn{7}{c}{~~} \\ [-1.8ex]
\multicolumn{7}{c}{SMC Clusters} \\ [0.5ex] \hline
\multicolumn{7}{c}{~~} \\ [-1.5ex] 
Kron 3 
 & 30   & 13.049\,$\pm$\,0.048 &                     & 0.943\,$\pm$\,0.063 &                     & 0.873\,$\pm$\,0.063 \\
 & 45   & 12.313\,$\pm$\,0.051 &                     & 0.994\,$\pm$\,0.066 &                     & 0.924\,$\pm$\,0.066 \\
 & 60   & 11.933\,$\pm$\,0.057 &                     & 0.978\,$\pm$\,0.074 &                     & 0.908\,$\pm$\,0.074 \\
 &\lhun & 11.411\,$\pm$\,0.085 &                     & 0.995\,$\pm$\,0.107 &                     & 0.925\,$\pm$\,0.107 \\ [0.5ex]
NGC 152 
 & 30   & 14.085\,$\pm$\,0.026 & 0.476\,$\pm$\,0.033 & 1.041\,$\pm$\,0.032 & 0.446\,$\pm$\,0.033 & 0.967\,$\pm$\,0.033 \\
 & 45   & 13.243\,$\pm$\,0.026 & 0.547\,$\pm$\,0.033 & 1.117\,$\pm$\,0.033 & 0.517\,$\pm$\,0.033 & 1.043\,$\pm$\,0.034 \\
 & 60   & 12.820\,$\pm$\,0.031 & 0.593\,$\pm$\,0.038 & 1.208\,$\pm$\,0.037 & 0.562\,$\pm$\,0.038 & 1.134\,$\pm$\,0.038 \\ 
 &\lhun & 12.327\,$\pm$\,0.053 & 0.570\,$\pm$\,0.064 & 1.168\,$\pm$\,0.063 & 0.540\,$\pm$\,0.064 & 1.094\,$\pm$\,0.064 \\ [0.5ex] 
NGC 265  
 & 30   & 12.735\,$\pm$\,0.034 & 0.259\,$\pm$\,0.069 & 0.567\,$\pm$\,0.078 & 0.227\,$\pm$\,0.069 & 0.490\,$\pm$\,0.078 \\
 & 45   & 12.299\,$\pm$\,0.048 & 0.336\,$\pm$\,0.097 & 0.737\,$\pm$\,0.101 & 0.304\,$\pm$\,0.097 & 0.660\,$\pm$\,0.101 \\
 & 60   & 12.176\,$\pm$\,0.074 & 0.333\,$\pm$\,0.152 & 0.721\,$\pm$\,0.160 & 0.301\,$\pm$\,0.152 & 0.643\,$\pm$\,0.160 \\
 &\lhun & 12.020\,$\pm$\,0.174 & 0.292\,$\pm$\,0.374 & 0.621\,$\pm$\,0.412 & 0.260\,$\pm$\,0.374 & 0.544\,$\pm$\,0.412 \\ [0.5ex]
NGC 339
 & 30   & 13.942\,$\pm$\,0.037 &                     & 0.912\,$\pm$\,0.052 &                     & 0.842\,$\pm$\,0.053 \\
 & 45   & 13.162\,$\pm$\,0.038 &                     & 0.961\,$\pm$\,0.052 &                     & 0.891\,$\pm$\,0.053 \\
 & 60   & 12.720\,$\pm$\,0.039 &                     & 0.975\,$\pm$\,0.053 &                     & 0.905\,$\pm$\,0.054 \\
 &\lhun & 12.124\,$\pm$\,0.044 &                     & 0.967\,$\pm$\,0.060 &                     & 0.897\,$\pm$\,0.061 \\ [0.5ex]
NGC 411
 & 30   & 12.944\,$\pm$\,0.035 & 0.597\,$\pm$\,0.050 & 0.946\,$\pm$\,0.049 & 0.570\,$\pm$\,0.050 & 0.880\,$\pm$\,0.049 \\
 & 45   & 12.484\,$\pm$\,0.039 & 0.586\,$\pm$\,0.054 & 0.957\,$\pm$\,0.052 & 0.559\,$\pm$\,0.054 & 0.891\,$\pm$\,0.052 \\
 & 60   & 12.191\,$\pm$\,0.044 & 0.562\,$\pm$\,0.061 & 0.914\,$\pm$\,0.059 & 0.535\,$\pm$\,0.061 & 0.848\,$\pm$\,0.059 \\
 &\lhun & 11.806\,$\pm$\,0.067 & 0.543\,$\pm$\,0.091 & 0.874\,$\pm$\,0.086 & 0.516\,$\pm$\,0.091 & 0.808\,$\pm$\,0.086 \\ [0.5ex]
NGC 419                                                                                                                       
 & 30   & 11.311\,$\pm$\,0.038 & 0.529\,$\pm$\,0.039 & 1.052\,$\pm$\,0.040 & 0.478\,$\pm$\,0.040 & 0.928\,$\pm$\,0.041 \\
 & 45   & 10.850\,$\pm$\,0.055 & 0.533\,$\pm$\,0.056 & 1.053\,$\pm$\,0.057 & 0.482\,$\pm$\,0.056 & 0.929\,$\pm$\,0.058 \\
 & 60   & 10.595\,$\pm$\,0.076 & 0.542\,$\pm$\,0.077 & 1.069\,$\pm$\,0.079 & 0.491\,$\pm$\,0.077 & 0.944\,$\pm$\,0.080 \\
 &\lhun & 10.304\,$\pm$\,0.161 & 0.571\,$\pm$\,0.162 & 1.097\,$\pm$\,0.166 & 0.520\,$\pm$\,0.162 & 0.972\,$\pm$\,0.166 \\ [0.5ex] \hline
\multicolumn{7}{c}{~~} \\ [-1.8ex]
\multicolumn{7}{c}{LMC Clusters} \\ [0.5ex] \hline
\multicolumn{7}{c}{~~} \\ [-1.5ex] 
NGC 1644
 & 30   & 13.184\,$\pm$\,0.012 & 0.407\,$\pm$\,0.016 & 0.811\,$\pm$\,0.020 & 0.363\,$\pm$\,0.017 & 0.702\,$\pm$\,0.022 \\
 & 45   & 12.922\,$\pm$\,0.016 & 0.417\,$\pm$\,0.020 & 0.838\,$\pm$\,0.028 & 0.373\,$\pm$\,0.020 & 0.729\,$\pm$\,0.029 \\
 & 60   & 12.824\,$\pm$\,0.022 & 0.416\,$\pm$\,0.027 & 0.835\,$\pm$\,0.043 & 0.371\,$\pm$\,0.027 & 0.726\,$\pm$\,0.043 \\
 &\lhun & 12.665\,$\pm$\,0.048 & 0.419\,$\pm$\,0.057 & 0.836\,$\pm$\,0.097 & 0.375\,$\pm$\,0.057 & 0.727\,$\pm$\,0.097 \\ [0.5ex]
NGC 1651
 & 30   & 13.332\,$\pm$\,0.047 &                     & 0.879\,$\pm$\,0.065 &                     & 0.743\,$\pm$\,0.068 \\
 & 45   & 12.769\,$\pm$\,0.055 &                     & 0.896\,$\pm$\,0.075 &                     & 0.760\,$\pm$\,0.078 \\
 & 60   & 12.506\,$\pm$\,0.069 &                     & 0.896\,$\pm$\,0.093 &                     & 0.760\,$\pm$\,0.096 \\
 &\lhun & 12.132\,$\pm$\,0.122 &                     & 1.147\,$\pm$\,0.151 &                     & 1.011\,$\pm$\,0.154 \\ [0.5ex]
NGC 1751
 & 30   & 12.911\,$\pm$\,0.039 & 0.599\,$\pm$\,0.055 & 1.324\,$\pm$\,0.045 & 0.517\,$\pm$\,0.055 & 1.126\,$\pm$\,0.046 \\
 & 45   & 12.378\,$\pm$\,0.051 & 0.636\,$\pm$\,0.072 & 1.349\,$\pm$\,0.059 & 0.554\,$\pm$\,0.072 & 1.151\,$\pm$\,0.059 \\
 & 60   & 12.088\,$\pm$\,0.069 & 0.652\,$\pm$\,0.096 & 1.416\,$\pm$\,0.077 & 0.570\,$\pm$\,0.096 & 1.218\,$\pm$\,0.077 \\
 &\lhun & 11.669\,$\pm$\,0.127 & 0.608\,$\pm$\,0.183 & 1.343\,$\pm$\,0.143 & 0.526\,$\pm$\,0.183 & 1.145\,$\pm$\,0.143 \\ [0.5ex]
NGC 1755
 & 30   & 10.239\,$\pm$\,0.022 & 0.216\,$\pm$\,0.024 & 0.504\,$\pm$\,0.026 & 0.130\,$\pm$\,0.025 & 0.295\,$\pm$\,0.028 \\
 & 45   &  9.969\,$\pm$\,0.031 & 0.202\,$\pm$\,0.033 & 0.482\,$\pm$\,0.035 & 0.115\,$\pm$\,0.034 & 0.272\,$\pm$\,0.036 \\
 & 60   &  9.859\,$\pm$\,0.047 & 0.193\,$\pm$\,0.048 & 0.470\,$\pm$\,0.049 & 0.106\,$\pm$\,0.048 & 0.260\,$\pm$\,0.050 \\
 &\lhun &  9.723\,$\pm$\,0.109 & 0.168\,$\pm$\,0.111 & 0.440\,$\pm$\,0.111 & 0.081\,$\pm$\,0.111 & 0.230\,$\pm$\,0.112 \\ [0.5ex]
NGC 1783  
 & 30   & 11.833\,$\pm$\,0.015 & 0.474\,$\pm$\,0.020 & 1.020\,$\pm$\,0.095 & 0.424\,$\pm$\,0.021 & 0.900\,$\pm$\,0.095 \\
 & 45   & 11.238\,$\pm$\,0.017 & 0.476\,$\pm$\,0.024 & 1.010\,$\pm$\,0.124 & 0.426\,$\pm$\,0.024 & 0.890\,$\pm$\,0.124 \\
 & 60   & 10.896\,$\pm$\,0.021 & 0.475\,$\pm$\,0.028 & 1.015\,$\pm$\,0.159 & 0.426\,$\pm$\,0.029 & 0.895\,$\pm$\,0.160 \\
 &\lhun & 10.393\,$\pm$\,0.034 & 0.485\,$\pm$\,0.045 & 1.078\,$\pm$\,0.262 & 0.435\,$\pm$\,0.045 & 0.958\,$\pm$\,0.262 \\ [0.5ex]
NGC 1806
 & 30   & 12.185\,$\pm$\,0.040 & 0.689\,$\pm$\,0.053 & 1.386\,$\pm$\,0.052 & 0.649\,$\pm$\,0.053 & 1.289\,$\pm$\,0.052 \\
 & 45   & 11.672\,$\pm$\,0.040 & 0.669\,$\pm$\,0.053 & 1.335\,$\pm$\,0.053 & 0.629\,$\pm$\,0.053 & 1.238\,$\pm$\,0.053 \\
 & 60   & 11.374\,$\pm$\,0.041 & 0.667\,$\pm$\,0.055 & 1.338\,$\pm$\,0.055 & 0.627\,$\pm$\,0.055 & 1.241\,$\pm$\,0.055 \\
 &\lhun & 10.999\,$\pm$\,0.046 & 0.669\,$\pm$\,0.063 & 1.344\,$\pm$\,0.065 & 0.629\,$\pm$\,0.063 & 1.247\,$\pm$\,0.065 \\ [0.5ex]
NGC 1831
 & 30   & 11.964\,$\pm$\,0.021 & 0.274\,$\pm$\,0.032 & 0.539\,$\pm$\,0.037 & 0.212\,$\pm$\,0.033 & 0.388\,$\pm$\,0.038 \\
 & 45   & 11.382\,$\pm$\,0.026 & 0.301\,$\pm$\,0.040 & 0.613\,$\pm$\,0.046 & 0.239\,$\pm$\,0.041 & 0.461\,$\pm$\,0.046 \\
 & 60   & 11.094\,$\pm$\,0.035 & 0.302\,$\pm$\,0.054 & 0.641\,$\pm$\,0.060 & 0.239\,$\pm$\,0.054 & 0.490\,$\pm$\,0.060 \\
 &\lhun & 10.729\,$\pm$\,0.067 & 0.288\,$\pm$\,0.104 & 0.630\,$\pm$\,0.117 & 0.225\,$\pm$\,0.104 & 0.479\,$\pm$\,0.117 \\
NGC 1846
 & 30   & 12.349\,$\pm$\,0.086 & 0.544\,$\pm$\,0.087 & 1.094\,$\pm$\,0.089 & 0.472\,$\pm$\,0.087 & 0.919\,$\pm$\,0.089 \\
 & 45   & 11.672\,$\pm$\,0.104 & 0.551\,$\pm$\,0.105 & 1.155\,$\pm$\,0.107 & 0.479\,$\pm$\,0.105 & 0.980\,$\pm$\,0.107 \\
 & 60   & 11.262\,$\pm$\,0.126 & 0.563\,$\pm$\,0.127 & 1.214\,$\pm$\,0.129 & 0.491\,$\pm$\,0.127 & 1.040\,$\pm$\,0.129 \\
 &\lhun & 10.675\,$\pm$\,0.203 & 0.532\,$\pm$\,0.205 & 1.166\,$\pm$\,0.208 & 0.460\,$\pm$\,0.205 & 0.992\,$\pm$\,0.209 
\\  [0.5ex] \hline
\multicolumn{6}{c}{~~} \\ [-1.8ex]
\end{tabular*}
\smallskip
\parbox{15cm}{
{\small
\noindent 
{\sl Note to Table \ref{t:photdata}:\/}~$^*$\,Radii are given in arcseconds.}}
\end{table*}

\addtocounter{table}{-1}
\begin{table*}
\small
\caption[ ]{(continued)} 
\begin{tabular*}{15cm}{@{\extracolsep{\fill}}lcrrrrr@{}} \hline \hline
\multicolumn{7}{c}{~~} \\ [-1.8ex]
Object & Aperture & \multicolumn{1}{c}{$V$} & \multicolumn{1}{c}{$V\!-\!R$} &
       \multicolumn{1}{c}{$V\!-\!I$} & \multicolumn{1}{c}{$(V\!-\!R)_0$} & 
       \multicolumn{1}{c}{$(V\!-\!I)_0$} \\
       & Radius\rlap{$^*$} & \multicolumn{1}{c}{[mag]} & \multicolumn{1}{c}{[mag]} & 
  \multicolumn{1}{c}{[mag]} & \multicolumn{1}{c}{[mag]} & 
  \multicolumn{1}{c}{[mag]} \\ [0.5ex] \hline
\multicolumn{7}{c}{~~} \\ [-1.8ex]
NGC 1866
 & 30   & 10.642\,$\pm$\,0.010 & 0.331\,$\pm$\,0.015 & 0.641\,$\pm$\,0.015 & 0.289\,$\pm$\,0.015 & 0.540\,$\pm$\,0.017 \\
 & 45   & 10.128\,$\pm$\,0.010 & 0.322\,$\pm$\,0.015 & 0.618\,$\pm$\,0.015 & 0.281\,$\pm$\,0.015 & 0.517\,$\pm$\,0.017 \\
 & 60   &  9.873\,$\pm$\,0.011 & 0.308\,$\pm$\,0.016 & 0.593\,$\pm$\,0.017 & 0.266\,$\pm$\,0.016 & 0.493\,$\pm$\,0.018 \\
 &\lhun &  9.533\,$\pm$\,0.013 & 0.319\,$\pm$\,0.021 & 0.650\,$\pm$\,0.022 & 0.278\,$\pm$\,0.021 & 0.549\,$\pm$\,0.023 \\ [0.5ex]
NGC 1868
 & 30   & 11.917\,$\pm$\,0.017 & 0.360\,$\pm$\,0.022 & 0.786\,$\pm$\,0.023 & 0.298\,$\pm$\,0.022 & 0.635\,$\pm$\,0.024 \\
 & 45   & 11.643\,$\pm$\,0.020 & 0.350\,$\pm$\,0.027 & 0.764\,$\pm$\,0.026 & 0.287\,$\pm$\,0.027 & 0.612\,$\pm$\,0.028 \\
 & 60   & 11.528\,$\pm$\,0.027 & 0.342\,$\pm$\,0.037 & 0.758\,$\pm$\,0.033 & 0.279\,$\pm$\,0.037 & 0.606\,$\pm$\,0.034 \\
 &\lhun & 11.382\,$\pm$\,0.056 & 0.328\,$\pm$\,0.080 & 0.760\,$\pm$\,0.065 & 0.266\,$\pm$\,0.080 & 0.608\,$\pm$\,0.066 \\ [0.5ex]
NGC 1978
 & 30   & 11.548\,$\pm$\,0.011 & 0.574\,$\pm$\,0.016 & 1.135\,$\pm$\,0.016 & 0.486\,$\pm$\,0.017 & 0.922\,$\pm$\,0.019 \\
 & 45   & 10.930\,$\pm$\,0.012 & 0.570\,$\pm$\,0.017 & 1.108\,$\pm$\,0.016 & 0.482\,$\pm$\,0.018 & 0.894\,$\pm$\,0.020 \\
 & 60   & 10.616\,$\pm$\,0.014 & 0.552\,$\pm$\,0.019 & 1.073\,$\pm$\,0.018 & 0.464\,$\pm$\,0.019 & 0.860\,$\pm$\,0.021 \\
 &\lhun & 10.203\,$\pm$\,0.020 & 0.557\,$\pm$\,0.027 & 1.088\,$\pm$\,0.025 & 0.469\,$\pm$\,0.028 & 0.875\,$\pm$\,0.028 \\ [0.5ex]
NGC 1987 
 & 30   & 12.626\,$\pm$\,0.021 & 0.422\,$\pm$\,0.028 & 0.964\,$\pm$\,0.032 & 0.369\,$\pm$\,0.029 & 0.836\,$\pm$\,0.033 \\
 & 45   & 12.249\,$\pm$\,0.031 & 0.412\,$\pm$\,0.042 & 0.912\,$\pm$\,0.049 & 0.359\,$\pm$\,0.042 & 0.784\,$\pm$\,0.049 \\
 & 60   & 12.000\,$\pm$\,0.042 & 0.456\,$\pm$\,0.056 & 1.026\,$\pm$\,0.063 & 0.403\,$\pm$\,0.056 & 0.897\,$\pm$\,0.064 \\
 &\lhun & 11.744\,$\pm$\,0.090 & 0.441\,$\pm$\,0.121 & 0.982\,$\pm$\,0.139 & 0.389\,$\pm$\,0.121 & 0.854\,$\pm$\,0.139 \\ [0.5ex]
NGC 2058  
 & 30   & 11.219\,$\pm$\,0.018 & 0.267\,$\pm$\,0.025 & 0.615\,$\pm$\,0.029 & 0.228\,$\pm$\,0.025 & 0.522\,$\pm$\,0.030 \\
 & 45   & 10.912\,$\pm$\,0.022 & 0.258\,$\pm$\,0.033 & 0.596\,$\pm$\,0.040 & 0.220\,$\pm$\,0.033 & 0.503\,$\pm$\,0.041 \\
 & 60   & 10.732\,$\pm$\,0.028 & 0.248\,$\pm$\,0.045 & 0.576\,$\pm$\,0.057 & 0.210\,$\pm$\,0.046 & 0.483\,$\pm$\,0.057 \\
 &\lhun & 10.451\,$\pm$\,0.053 & 0.251\,$\pm$\,0.091 & 0.639\,$\pm$\,0.110 & 0.213\,$\pm$\,0.091 & 0.546\,$\pm$\,0.110 \\ [0.5ex]
NGC 2134
 & 30   & 11.521\,$\pm$\,0.011 & 0.264\,$\pm$\,0.016 & 0.572\,$\pm$\,0.017 & 0.185\,$\pm$\,0.016 & 0.382\,$\pm$\,0.018 \\
 & 45   & 11.158\,$\pm$\,0.013 & 0.263\,$\pm$\,0.018 & 0.572\,$\pm$\,0.020 & 0.185\,$\pm$\,0.019 & 0.382\,$\pm$\,0.021 \\
 & 60   & 10.972\,$\pm$\,0.017 & 0.262\,$\pm$\,0.023 & 0.563\,$\pm$\,0.025 & 0.184\,$\pm$\,0.023 & 0.373\,$\pm$\,0.026 \\
 &\lhun & 10.742\,$\pm$\,0.031 & 0.295\,$\pm$\,0.042 & 0.663\,$\pm$\,0.047 & 0.217\,$\pm$\,0.042 & 0.473\,$\pm$\,0.047 \\ [0.5ex]
NGC 2136
 & 30   & 11.059\,$\pm$\,0.012 & 0.327\,$\pm$\,0.016 & 0.728\,$\pm$\,0.016 & 0.254\,$\pm$\,0.016 & 0.549\,$\pm$\,0.018 \\
 & 45   & 10.715\,$\pm$\,0.015 & 0.308\,$\pm$\,0.019 & 0.683\,$\pm$\,0.019 & 0.235\,$\pm$\,0.019 & 0.505\,$\pm$\,0.020 \\
 & 60   & 10.500\,$\pm$\,0.019 & 0.287\,$\pm$\,0.023 & 0.642\,$\pm$\,0.023 & 0.213\,$\pm$\,0.023 & 0.463\,$\pm$\,0.024 \\
 &\lhun & 10.295\,$\pm$\,0.040 & 0.311\,$\pm$\,0.044 & 0.684\,$\pm$\,0.044 & 0.237\,$\pm$\,0.044 & 0.505\,$\pm$\,0.044 \\ [0.5ex]
NGC 2154
 & 30   & 12.810\,$\pm$\,0.033 & 0.473\,$\pm$\,0.044 & 1.005\,$\pm$\,0.054 & 0.419\,$\pm$\,0.044 & 0.873\,$\pm$\,0.055 \\
 & 45   & 12.329\,$\pm$\,0.046 & 0.489\,$\pm$\,0.060 & 1.067\,$\pm$\,0.075 & 0.435\,$\pm$\,0.061 & 0.935\,$\pm$\,0.075 \\
 & 60   & 12.099\,$\pm$\,0.064 & 0.488\,$\pm$\,0.085 & 1.060\,$\pm$\,0.105 & 0.433\,$\pm$\,0.085 & 0.928\,$\pm$\,0.106 \\
 &\lhun & 11.851\,$\pm$\,0.138 & 0.494\,$\pm$\,0.184 & 1.137\,$\pm$\,0.229 & 0.439\,$\pm$\,0.184 & 1.005\,$\pm$\,0.229 \\ [0.5ex]
NGC 2155
 & 30   & 13.417\,$\pm$\,0.094 & 0.556\,$\pm$\,0.104 & 1.007\,$\pm$\,0.157 & 0.500\,$\pm$\,0.104 & 0.872\,$\pm$\,0.157 \\
 & 45   & 12.909\,$\pm$\,0.132 & 0.581\,$\pm$\,0.145 & 1.030\,$\pm$\,0.217 & 0.525\,$\pm$\,0.145 & 0.894\,$\pm$\,0.218 \\
 & 60   & 12.696\,$\pm$\,0.193 & 0.601\,$\pm$\,0.211 & 1.006\,$\pm$\,0.322 & 0.545\,$\pm$\,0.211 & 0.871\,$\pm$\,0.322 \\
 &\lhun & 12.593\,$\pm$\,0.485 & 0.680\,$\pm$\,0.524 & 0.906\,$\pm$\,0.863 & 0.624\,$\pm$\,0.524 & 0.770\,$\pm$\,0.863 \\ [0.5ex]
NGC 2162
 & 30   & 13.284\,$\pm$\,0.045 & 0.535\,$\pm$\,0.051 & 1.087\,$\pm$\,0.055 & 0.472\,$\pm$\,0.051 & 0.935\,$\pm$\,0.043 \\
 & 45   & 12.846\,$\pm$\,0.065 & 0.563\,$\pm$\,0.073 & 1.122\,$\pm$\,0.078 & 0.501\,$\pm$\,0.073 & 0.971\,$\pm$\,0.058 \\
 & 60   & 12.613\,$\pm$\,0.092 & 0.535\,$\pm$\,0.103 & 1.064\,$\pm$\,0.112 & 0.472\,$\pm$\,0.104 & 0.912\,$\pm$\,0.081 \\
 &\lhun & 12.350\,$\pm$\,0.198 & 0.563\,$\pm$\,0.221 & 1.087\,$\pm$\,0.240 & 0.500\,$\pm$\,0.221 & 0.936\,$\pm$\,0.167 \\ [0.5ex]
NGC 2164
 & 30   & 10.774\,$\pm$\,0.033 &                     & 0.312\,$\pm$\,0.046 &                     & 0.153\,$\pm$\,0.053 \\
 & 45   & 10.490\,$\pm$\,0.033 &                     & 0.332\,$\pm$\,0.046 &                     & 0.173\,$\pm$\,0.053 \\
 & 60   & 10.309\,$\pm$\,0.033 &                     & 0.370\,$\pm$\,0.046 &                     & 0.211\,$\pm$\,0.053 \\
 &\lhun & 10.105\,$\pm$\,0.033 &                     & 0.386\,$\pm$\,0.047 &                     & 0.227\,$\pm$\,0.054 \\ [0.5ex]
NGC 2173
 & 30   & 12.999\,$\pm$\,0.037 & 0.573\,$\pm$\,0.064 & 1.269\,$\pm$\,0.055 & 0.511\,$\pm$\,0.064 & 1.117\,$\pm$\,0.056 \\        
 & 45   & 12.481\,$\pm$\,0.050 & 0.573\,$\pm$\,0.087 & 1.230\,$\pm$\,0.073 & 0.510\,$\pm$\,0.088 & 1.078\,$\pm$\,0.074 \\        
 & 60   & 12.306\,$\pm$\,0.074 & 0.558\,$\pm$\,0.131 & 1.218\,$\pm$\,0.108 & 0.496\,$\pm$\,0.131 & 1.066\,$\pm$\,0.109 \\
 &\lhun & 12.007\,$\pm$\,0.224 & 0.540\,$\pm$\,0.277 & 1.207\,$\pm$\,0.225 & 0.477\,$\pm$\,0.277 & 1.056\,$\pm$\,0.225 \\ [0.5ex]
NGC 2213
 & 30   & 12.975\,$\pm$\,0.019 & 0.575\,$\pm$\,0.023 & 1.155\,$\pm$\,0.023 & 0.504\,$\pm$\,0.024 & 0.984\,$\pm$\,0.025 \\
 & 45   & 12.585\,$\pm$\,0.027 & 0.600\,$\pm$\,0.032 & 1.184\,$\pm$\,0.032 & 0.530\,$\pm$\,0.032 & 1.013\,$\pm$\,0.033 \\
 & 60   & 12.477\,$\pm$\,0.042 & 0.596\,$\pm$\,0.049 & 1.170\,$\pm$\,0.048 & 0.526\,$\pm$\,0.049 & 0.999\,$\pm$\,0.049 \\
 &\lhun & 12.365\,$\pm$\,0.102 & 0.640\,$\pm$\,0.116 & 1.225\,$\pm$\,0.115 & 0.570\,$\pm$\,0.116 & 1.054\,$\pm$\,0.115 \\ [0.5ex]
NGC 2231
 & 30   & 13.648\,$\pm$\,0.016 & 0.562\,$\pm$\,0.026 & 1.073\,$\pm$\,0.090 & 0.500\,$\pm$\,0.026 & 0.921\,$\pm$\,0.090 \\
 & 45   & 13.231\,$\pm$\,0.020 & 0.541\,$\pm$\,0.034 & 1.002\,$\pm$\,0.139 & 0.479\,$\pm$\,0.034 & 0.851\,$\pm$\,0.139 \\
 & 60   & 13.206\,$\pm$\,0.021 & 0.539\,$\pm$\,0.034 & 0.995\,$\pm$\,0.143 & 0.477\,$\pm$\,0.034 & 0.844\,$\pm$\,0.143 \\
 &\lhun & 12.655\,$\pm$\,0.055 & 0.517\,$\pm$\,0.095 & 0.779\,$\pm$\,0.517 & 0.455\,$\pm$\,0.095 & 0.627\,$\pm$\,0.517 
 \\ [0.5ex] \hline
\multicolumn{6}{c}{~~} \\ [-1.8ex]
\end{tabular*}
\smallskip
\parbox{15cm}{
%\baselineskip=0.98\normalbaselineskip
{\small
\noindent 
{\sl Note to Table:\/}~$^*$\,Radii are given in arcseconds.}}
\end{table*}

\end{document}